\documentstyle[12pt,aaspp4]{article} 
\def\tempest%
{\begin{array}{ccc} 
1 & 1 & 1 \\ 
1 & 1 & 1 \\ 
4 & 3 & 8 
\end{array}}

\def\kms{{\rm km}\,{\rm s}^{-1}}
\def\br{{\bf r}}
\begin{document}

\title{Selection of Nearby Microlensing Candidates for Observation by
SIM}
\author 
{Andrew Gould}
\affil{Ohio State University, Department of Astronomy, Columbus, OH 43210} 
\affil{E-mail: gould@astronomy.ohio-state.edu} 
\begin{abstract} 

I investigate the prospects for using the Space Interferometry Mission (SIM) 
to measure the masses of nearby
stars from their astrometric deflection of more distant sources, as 
originally suggested by Paczy\'nski and by Miralda-Escud\'e.  I derive an
analytic expression for the total observing time $T_{\rm tot}$ required
to measure the masses of a fixed number of stars to a given precision.
I find that $T_{\rm tot}\propto r_{\rm max}^{-2}$, where $r_{\rm max}$ is
the maximum radius to which candidates are searched,
or $T_{\rm tot}\propto \mu_{\rm min}^2$, where $\mu_{\rm min}$ is
the minimum proper motion to which candidates are searched.  I show that
$T_{\rm tot}$ can be reduced by a factor 4 if source availability is
extended from $V_s=17$ to $V_s=19$.  Increasing $r_{\rm max}$ and $V_s$
and decreasing $\mu_{\rm min}$
all require a significantly more agressive approach to finding candidates.
A search for candidates can begin by making use of the Luyton
proper motion catalog together with the USNO-A2.0 all-sky astrometric
catalog.  However, a thorough search would require the
all-sky USNO-B proper-motion catalog which is not yet available.  
The follow-up observations
necessary to prepare for the mission will become more difficult the longer
they are delayed because the candidate pairs are typically already within 
$1''$ and are getting closer.

\end{abstract} 
\newpage

\section{Introduction} 

	Refsdal (1964) pointed out that it should be possible to measure
the masses of nearby field stars from the astrometric deviation
they induce on more distant sources as they pass by the latter.
To be practical, the two stars must pass within ${\cal O}(1'')$ of each other.
Paczy\'nski (1995,1998) and Miralde-Escud\'e (1996) examined this idea in the
context of the current rapid improvements in astrometric capability.  
They made rough estimates of the number of mass measurements that could be 
obtained using various ground-based and space-based facilities.  Here
I re-examine this problem specifically guided by the capabilities and
requirements of the Space Interferometry Mission (SIM).

	The planned SIM launch date is 2005 and the minimum mission lifetime
is 5 years.  In order to carry out mass measurements, two steps must
be completed prior to launch.  First, one must identifify candidate pairs of
stars from a proper-motion catalog:  a nearby ``lens'' star
must be found that is likely to pass sufficiently close to a more 
distant ``source'' star to cause a large deflection of light and so permit
a precise measurement of this deflection.  
Second, given the quality of the catalogs that will be available
in the near future, it will generally not be possible to predict which of
the candidates will be the best to make precise mass measurements with 
a modest amount of observing time.  Rather, it will be necessary to perform
follow-up observations of these candidates prior to the event in order to 
determine the impact parameter (angular separation $\beta$ at the point
of closest approach).  Typically, the candidates are {\it already}
closer than $1''$, often much closer.  Moreover, in many cases, one star is
substantially brighter than the other.  Hence, the follow-up observations
will usually require adaptive optics or the Hubble Space Telescope.  These
requirements will grow more severe as time passes.  In brief, 
preparation for mass measurements using SIM should proceed without delay.

	Because SIM observing time comes at a high premium, my approach 
is to rank candidates by the amount of observing time that is required to make 
a mass measurement of fixed precision.  I then use this framework to
characterize and evaluate various selection strategies.  

	The probability that it is possible to measure the mass
of a given foreground star grows monotonically with its proper motion
and is linear in the proper motion in most cases.  Hence, a survey
based on an ideal star catalog (not affected by magnitude limits or 
crowding) would investigate foreground stars down to some minimum
proper motion $\mu_{\rm min}$.  On the other hand, for stars of sufficiently 
low luminosity, the magnitude limits of the underlying catalog will impose
an effective distance limit, $r_{\rm max}$.  Thus, it is important to
consider both forms of selection.  In practice, the actual selection process
may also be affected by crowding, but I will not consider crowding
explicitly in this paper.  Rather, one may think of crowding as imposing
an indirect constraint on $\mu_{\rm min}$ or $r_{\rm max}$.

	I derive simple expressions for the total observing time 
$T_{\rm tot}$ needed to make ${\cal N}$ mass measurements.  
 For fixed ${\cal N}$, I show that $T_{\rm tot}\propto r_{\rm max}^{-2}$
for distance limited surveys, and $T_{\rm tot}\propto \mu_{\rm min}^{2}$
for proper-motion limited surveys.  Hence, minimization of the observing time
requires pushing $r_{\rm max}$ out as far as possible or pushing
$\mu_{\rm min}$ as low as possible.  The sample of candidates will then
have on average smaller proper motions meaning that
they are even closer on the sky today, thus making
follow-up observations even more difficult.  In addition,
I show that by extending the available sources from $V_s=17$ to $V_s=19$,
one can decrease $T_{\rm tot}$ by a factor of 4 despite the lower flux
from these fainter sources.  However, to determine which of these fainter
sources are really usable requires a much more precise estimate of their
expected impact parameters, that is, even more precise measurements of
their current positions despite the larger disparity in the source/lens
flux ratio.  Obviously these measurements will also become more difficult
with time.

	To carry out a search for candidates it would be best to begin
with an all-sky proper motion catalog.  Such a catalog is currently
being prepared by the US Naval Observatory (D.\ Monet 1999, private
communication) but has not yet been released.  In the meantime, one can make
a good start using the Luyton (1979) proper motion catalog in combination
with the USNO-A2.0 astrometric catalog (Monet 1998).  I briefly describe how
to carry out such a search.

\section{Required Observation Time For an Individual Lens}

	Consider a nearby star (``the lens'') of mass $M$ and 
at distance $r$ that passes
within an angle $\beta$ of a more distant star (``the source'') at $r_s$.  
The source light will then be deflected by 
an angle $\alpha = 4 G M/(\beta r c^2)$
at the point of closest approach.  Consequently, the source will appear
displaced by $\tilde \alpha\equiv
\alpha(1 - r/r_s)= 4 G M\pi_{\rm rel}/({\rm AU}\beta c^2)$ 
relative to the position expected in the absence of lensing.  
One can therefore determine the mass of the lens by measuring this 
displacement, provided that $\beta$ and the relative parallax $\pi_{\rm rel}$ 
are known.  Note that
\begin{equation}
\alpha = 80\,\mu{\rm as}
{M\over M_\odot}\,\biggl({\beta r\over 100\,{\rm AU}}\biggr)^{-1}.
\label{eqn:alphaeval}
\end{equation}

	Assuming photon-limited astrometry measurements, the total amount of 
observing time $\tau$ required to achieve a fixed fractional error in the mass 
measurement then depends on three factors.  First, the measurement is
easier the bigger $\tilde\alpha$: the factional error for fixed observing
time falls as $\tilde \alpha^{-1}$ and so for fixed fractional error,
$\tau \propto \tilde \alpha^{-2}$. Second,
the measurement is easier the brighter the source magnitude $V_s$,
$\tau\propto 10^{0.4 V_s}$.  Third, the time required depends on the 
geometry of the encounter.  The geometry can be described in terms of
the angular coordinates $(\beta,\lambda)$ of the source-lens 
separation vector at the time ($t=0$) of the midpoint of the mission
and the angular displacement $\mu\Delta t$ of the lens relative to the 
source during the course of the mission.  Here $\beta$ is the source-lens
separation at the time of closest approach ($t=t_0$), $\mu$ is the relative
source-lens proper motion, $\Delta t$ is the duration of the mission, and
$\lambda = -\mu t_0$.  I therefore write,
\begin{equation}
\tau = T_0 \biggl({\tilde\alpha\over \alpha_0}\biggr)^{-2}\,
10^{0.4(V_s-17)}\,\gamma\biggl({\lambda
\over \beta},{\mu \Delta t\over \beta}
\biggr),
\label{eqn:taudef}
\end{equation}
where $\gamma$ is a function to be described below, and where
$\alpha_0$ and $T_0$ are convenient normalization factors.  For definiteness,
I will take the required mass precission to be $\sigma_M/M=1\%$ 
and will arbitrarily
adopt $\alpha_0=100\,\mu$as.  I will normalize $\gamma$ so that it is
effectively the number of equal-duration measurements that must be made.
I characterize SIM astrometry as requiring 1 minute to achieve 40 $\mu$as
precision in 1 dimension at $V_s=17$.  Then 
\begin{equation}
T_0 = \biggl({\sigma_M\over M}\biggr)^{-2}\biggr({40\,\mu\rm as\over \alpha_0}
\biggr)^2\,{\rm min} = 27\,{\rm hours}.
\label{eqn:tzerodef}
\end{equation}

	To estimate $\gamma$, I consider sets of observations over the
angular interval $[\lambda_-,\lambda_+]$ where $\lambda_\pm=
\lambda\pm \mu\Delta t/2$, and solve simulteously for six source parameters:
the two-dimensional angular position at the midpoint, the two-dimensional
proper motion, the parallax, and $\tilde\alpha$.  Even though very little
information can be obtained about $\tilde\alpha$ from astrometry measurements
parallel to its direction of motion, I include such measurements in order
to be sensitive to other kinds of apparent source acceleration (e.g.\
gravitational).  Without such a check, the mass measurement could not
be considered reliable.  I then optimize these observations for the
measurement of $\tilde\alpha$.  Generally, the optimum configuration has 
roughly equal total exposure times at the point of closest approach and
near the beginning and end of the experiment, and has no observations at
other times.  I take $\Delta t$ to be
5 years, but the results would be the same for any value provided that
$\Delta t \gg 1\,$yr.  I find that $\gamma$ achieves a minimum, $\gamma\sim 10$
when $\lambda_-<-2\beta$ and $\lambda_+>+2\beta$. For example,
$\gamma(0,x)=10$ for $x\geq 4$. Some other indicative
values are $\gamma(0,2) = 21$, $\gamma(0,1)=99$, $\gamma(0.75,2.5)=24$,
$\gamma(0.25,1.5)=39$.  Thus, if the source does not move by a distance at
least equal to the impact parameter on either side of the lens during the
course of the observations, $\gamma$ becomes very high.  In my analysis
below, I will incorporate the exact values of $\gamma$ for each configuration.
But qualitatively one can think of $\gamma$ as being 
\begin{equation}
\biggl[\gamma\biggl({\lambda\over \beta},{\mu\Delta t\over \beta}\biggr)
\biggr]^{-1} \sim \gamma_*^{-1}\,
\Theta\biggl(\lambda - \beta + {\mu\Delta t\over 2}\biggr)
\Theta\biggl(-\lambda - \beta + {\mu\Delta t\over2}\biggr),
\label{eqn:gammaapprox}
\end{equation}
where $\Theta$ is a step function, and $\gamma_*=10$.

\section{Observing-time Distribution}

	From the previous section, a star with $M=M_\odot$, $r=100\,$pc,
$\beta=1''$, and a minimal $\gamma$ would require about 420 hours of
observation time for a 1\% mass measurement.  Hence, it is prudent to
consider how one might find pairs of stars with the most favorable 
characteristics.  I begin by writing down the observing-time distribution for
an arbitrarily selected sample but with lenses of fixed mass $M$,
\begin{equation}
{d{\cal N}\over d T} 
= \int d^3 r\, d^3 r_s d V_s d v_\perp n(\br) n_s(V_s,\br_s)
f(\br,v_\perp) S(\br,\br_s,v_\perp,V_s, ...)
\delta[T -\tau(V_s,M,b,\ell,v_\perp \Delta t, r,r_s)].
\label{eqn:qoft}
\end{equation}
Here $n(\br)$ is the number density of lenses as a function of their
position, $n_s(V_s,\br_s)$ is the number density of sources as a function
of their magnitude and position, $v_\perp=r\mu$ is the transverse
speed of the lens relative to the observer-source line of sight, 
$f(\br,v_\perp)$ is the transverse speed distribution as a function of
position, $S$ is the selection function (with possibly many variables
in addition to those explicitly shown), $b = r\beta$, $\ell = r\lambda$,
and $\delta$ is a Dirac delta
function.  Equation (\ref{eqn:qoft}) cannot be simplified without additional
assumptions.  As I introduce these assumptions, I will briefly outline
their impact.  Some of the simplifications will then be discussed in greater
detail below.

	I first assume that $r\ll r_s$.  This is an excellent approximation
for disk lenses although it is not as good for halo lenses.  It has
two simplifying effects: $\tilde \alpha \rightarrow \alpha$,
so $\tau = \tau(V_s,M,b,\ell,v_\perp\Delta t)$, and the 3-space
density of sources $n_s$ can be replaced with the projected surface density
$\phi(V_s,\Omega)$, where $\Omega$ is position on the sky.  (To be more
precise $\phi$ is the density of sources in the neighborhood of the
{\it lens} postion $\Omega$.)
Second, I assume that the product of the selection 
function and number density can be written,
\begin{equation}
n(\br)S(\br,\br_s,v_\perp,V_s, ...) \rightarrow n \Theta(r_{\rm max} - r),
\label{eqn:selectsim}
\end{equation}
where $n$ is now assumed to be uniform, and $r_{\rm max}$ is a maximum
search radius.  In fact, this is an oversimplification.  A major
focus of the present study is to determine what effect the selection
function has on the observing-time distribution.  The best way to do this
is to begin with this simplified picture.  With these assumptions, equation
(\ref{eqn:qoft}) can be written,
\begin{equation}
{d{\cal N}\over d T}  = \int d v_\perp\, d V_s
\int d\beta\,d\lambda\, d r\,r^2 
n \Theta(r_{\rm max}-r) \delta(T-\tau) \int d \Omega\,\phi(V_s,\Omega) 
f(v_\perp,\br).
\label{eqn:qoft2}
\end{equation}
In this form, the
integration still cannot be factored because of the correlation between the
speed distribution of the lenses and the distribution of sources.  I
therefore assume that $f(v_\perp,\br)\rightarrow f(v_\perp)$, i.e.,
that the speed distribution does not depend on position.  This is actually
a very minor assumption, provided that the speed distribution is taken to
be the average over the Galactic plane where the majority of the source
stars are.  Two of the integrals can then be evaluated directly, and
equation (\ref{eqn:qoft2}) becomes,
\begin{equation}
{d{\cal N}\over d T}  = n r_{\rm max}
\int d b\,d\ell\, d V_s\, d v_\perp\, f(v_\perp)\phi(V_s)\,
\delta [T-\tau(V_s,M,b,\ell,v_\perp\Delta t)],
\label{eqn:qoft3}
\end{equation}
where $\phi(V_s)\equiv \int d\Omega \phi(V_s,\Omega)$ is the 
luminosity function integrated over the entire sky.  Equation (\ref{eqn:qoft3})
already contains an important result: the number of lenses available for
measurement at fixed observing time is directly proportional to $r_{\rm max}$,
the physical depth to which they are searched.

	To further evaluate the integral, first define,
\begin{equation}
G(\gamma';x) = \int d y\, \delta[\gamma' - \gamma(y,x)].
\label{eqn:Gdef}
\end{equation}
Then the integral can be written
\begin{equation}
{d{\cal N}\over d T}  
= n r_{\rm max} \int d\Gamma d V_s \phi(V_s) \delta [T-\tau(V_s,\Gamma)]
H(\Gamma),
\label{eqn:qoft4}
\end{equation}
where
\begin{equation}
H(\Gamma) = \int d \gamma \, d b\, d v_\perp\, b f(v_\perp) 
G\biggl(\gamma;{v_\perp \Delta t\over b}\biggr)
\delta\biggl[\Gamma - \biggl({4 G M \over \alpha_0 b c^2}\biggr)^{-2}\gamma
\biggr].
\label{eqn:Hdef}
\end{equation}

\subsection{Analytic Estimate}

	I will evaluate equation (\ref{eqn:qoft4}) explicitly in \S\ 3.2 
below.  However, it is also instructive to make an analytic estimate of
this equation with the help of a few approximations.  First,  I assume
that all the sources have the same magnitude, $\phi(V_s) = N\delta(V_s-17)$,
where $N$ is the total number of source stars. Hence,
\begin{equation}
{d{\cal N}\over d T}  = {n N r_{\rm max}\over T_0}H\biggl({T\over T_0}\biggr).
\label{eqn:qoft5}
\end{equation}
Second, I use the approximation
(\ref{eqn:gammaapprox}) to estimate $G$,
\begin{equation}
G(\gamma;x) =(x-2)\Theta(x-2)\delta(\gamma_*-\gamma),\qquad (\gamma_*= 10).
\label{eqn:gammaapprox2}
\end{equation}
Third, I take 
$f(v_\perp)=\delta(v_\perp -v_{*})$, where $v_{*}$ is a typical transverse
speed for the lens population.  
Then
\begin{equation}
H(\Gamma) = \Gamma^{-1/2}b_0\gamma_*^{-1/2}
\biggl[{v_{*}\Delta t} - 4 b_0
\biggl({\Gamma\over\gamma_*}\biggr)^{1/2}\biggr],\qquad 
b_0\equiv {2G M\over \alpha_0 c^2},
\label{eqn:Heval}
\end{equation}
where I have suppressed the $\Theta$ function that limits the range
of validity to $\Gamma < \gamma_*(\alpha_0 c^2 v_{*}\Delta t/8G M)^2$.
Combining equations (\ref{eqn:qoft5}) and (\ref{eqn:Heval}), I obtain,
\begin{equation}
{d{\cal N}\over d T}  = {2 G \rho N r_{\rm max}v_*\Delta t\over 
(\gamma_* T T_0)^{1/2}\alpha_0 c^2},
\qquad \biggl[T\ll 
T_0 \gamma_*\biggl({\alpha_0 c^2 v_{*}\Delta t\over 8G M}\biggr)^2\biggr],
\label{eqn:qoft6}
\end{equation}
where $\rho\equiv n M$.  The limiting condition in equation (\ref{eqn:qoft6})
comes from assuming that the first term in brackets in equation 
(\ref{eqn:Heval}) is much greater than the second.
Equation (\ref{eqn:qoft6}) tells us that the observing-time distribution
depends on the type of lens only through its mass density $\rho$, its
typical velocity $v_*$, and
the cutoff which scales as $(v_*/M)^{2}$.

	A sensible observing strategy will naturally focus on the lenses
that require the least observing time.
I therefore consider a program
that measures the masses of all lenses requiring observing times less
than some maximum, $T_{\rm max}$.  The total observing time 
$T_{\rm tot}$ can 
then be expressed as a function of the total number of stars 
observed, ${\cal N}$, and of the other parameters:
\begin{equation}
T_{\rm tot} = \int_0^{T_{\rm max}} d T\, T{d{\cal N}\over d T},\qquad
{\cal N} = \int_0^{T_{\rm max}} d T\, {d{\cal N}\over d T},
\label{eqn:tandn}
\end{equation}
\begin{equation}
T_{\rm tot} = {1\over 3}{\cal N}^3
\biggl({4 G \rho N r_{\rm max} v_*\Delta t\over \alpha_0 c^2}\biggr)^{-2}
\gamma_* T_0.
\label{eqn:totaltime}
\end{equation}

	Equation (\ref{eqn:totaltime}) is one of the major results of this
paper.  It states that the total observing time required to 
measure the masses of a fixed number of lenses scales inversely as the
square of the search radius $r_{\rm max}$ of the sample.  Given the
premium on SIM time, this result implies that the search should be pushed
to as large a radius as possible.  I discuss the prospects for doing this
in \S\ 4.

	The total time can be written out explicitly
\begin{equation}
T_{\rm tot} = 230\,{\rm hours}\, 
\biggl({{\cal N}\over 5}\biggr)^3
\biggl({\rho\over 0.01\,M_\odot\,\rm pc^{-3}}\biggr)^{-2}
\biggl({v_*\Delta t\over 35\,\rm AU}\biggr)^{-2}
\biggl({r_{\rm max}\over 100\,\rm pc}\biggr)^{-2}
\biggl({N\over 10^8}\biggr)^{-2},
\label{eqn:totaltime2}
\end{equation}
where I have assumed a mission lifetime of $\Delta t=5\,$yrs and
normalized the transverse speed to a typical disk value
$v_*=33\,\kms$ and 
the density to approximately 1/3 of the local stellar disk density
(Gould, Bahcall, \& Flynn 1997).  That is I consider that one is interested
in one (or perhaps several) subsets of the whole disk population.
I have also assumed a total of $N=10^8$ stars at $V_s=17$ over the whole sky
(Mihalas \& Binney 1981).
This estimate incoporates a maximum observing time per object,
\begin{equation}
T_{\rm max} = {3\over \cal N}T_{\rm tot},
\label{eqn:tmax}
\end{equation}
which must be well under the cutoff in equation (\ref{eqn:qoft6}) given by
\begin{equation}
T_{\rm cut} = 13\,{\rm hours}\,\biggl({v_*\Delta t\over 35\,\rm AU}\biggr)^2
\biggl({M\over M_\odot}\biggr)^{-2}.
\label{eqn:tcut}
\end{equation}
If $T_{\rm cut}\ga T_{\rm max}$, then the scaling relation 
(\ref{eqn:totaltime}) $({\cal N}\propto T_{\rm tot}^{1/3})$ is no
longer satisfied.  See Figure \ref{fig:one}, below.
The cutoff is satisfied for low mass disk stars
(assuming only 5 mass measurements are desired) but becomes more difficult
for higher masses.  

	Another important feature of equation (\ref{eqn:totaltime}) is
that $T_{\rm tot}\propto 10^{0.4(V_s-17)}T_0/N^2$.  
Thus, if we compare $V_s=17$ and
$V_s=18$, the latter are 2.5 times fainter and so require 2.5 times greater
$10^{0.4(V_s-17)}T_0$, the observing time for a single astrometric measurement
of precision $\alpha_0$.  
On the other hand, there are approximately 1.9 times as many stars
(Milhalas \& Binney 1981) and so $T_{\rm tot}$ is actually smaller by
a factor $\sim 0.7$.  It should be noted, however, that the shorter
observing time comes about because the impact parameter $b$ is typically
1.9 times smaller.  In \S\ 4, I will discuss the prospects for recognizing
when such close encounters will occur.

\subsection{Numerical Estimates}

	To test the estimates derived in \S\ 3.1,  I continue to approximate
$\phi$ as a $\delta$ function, but otherwise carry out the full integration
indicated by equations 
(\ref{eqn:qoft4}) and (\ref{eqn:Hdef}).  I take
the velocity distribution to be a two-dimensional Gaussian with 
(one-dimensional) dispersion typical of foreground
objects in the Galactic plane:
$\sigma^2 = \sigma_U^2/4 + \sigma_V^2/4 + \sigma_W^2/2$, where
$\sigma_U=34\,\kms$, $\sigma_V=28\,\kms$, and $\sigma_W=20\,\kms$.
This yields $\sigma=26\,\kms$ for which the mean speed is
$v_*= (\pi/2)^{1/2}\sigma = 33\,\kms$ (as used in \S\ 3.1).
Figure \ref{fig:one} shows the results for $M=M_\odot$ {(\it bold curve)}
and $M=0.1\,M_\odot$ {(\it solid curve)}.  The agreement with the
analytic prediction from equation (\ref{eqn:totaltime2}) ({\it dashed line})
is excellent.  Equations (\ref{eqn:tmax}) and (\ref{eqn:tcut}) predict
that the cutoff should be at ${\cal N}\sim 1.5 (M/M_\odot)^{-1}$, or
at ${\cal N}\sim 1.5$ and ${\cal N}\sim 15$ for the two cases shown.
In fact the actual values are about 2.5 times higher.  Most of this difference
(a factor of 2) is due to the fact that the velocity distribution is not
a $\delta$ function, and the higher-speed stars are more likely to be
candidates and are less affected by the threshold.  

	Figure \ref{fig:two} shows the same quantities for six different
magnitude bins of the luminosity function (Mihalas \& Binney 1981), and
$M=M_\odot$.
The $V_s=17$ curve (same as in Fig. \ref{fig:one}) is shown as
bold dashed line, and the others $V_s=14,15,16,18,19$ are shown as solid
lines.  The curves can be separately identified by noting that the
cutoff increases with magnitude.
The upper bold line shows the result of combining all of these while
the lower bold line shows the result of combining the four bins with 
$V_s\leq 17$ together.  Note that each of three bins $V_s=17,18,19$
contribute about equally
(for $T_{\rm tot}\la 100\,$hours).  This is because the longer integration
times required for the fainter sources are compensated by the fact that
they are more numerous and hence closer on average to the lenses.  

\subsection{Proper-motion selection}

	As I discussed in the introduction, in some regimes the selection
function is best described as a cut on distance and in other it is best
described as a cut on proper motion.  So far, I have focused on
selection by distance.  See equation (\ref{eqn:selectsim}).  Had I instead
selected on proper motion,
\begin{equation}
n(\br)S(\br,\br_s,v_\perp,V_s, ...) \rightarrow n 
\Theta \biggl({v_\perp\over r}-\mu_{\rm min}\biggr),
\label{eqn:selectsim2}
\end{equation}
then equations (\ref{eqn:qoft4}) and (\ref{eqn:Hdef}) would be replaced by
\begin{equation}
{d{\cal N}\over d T}  
= n \int d r d\Gamma d V_s \phi(V_s) \delta [T-\tau(V_s,\Gamma)]
H(\Gamma;r\mu_{\rm min}),
\label{eqn:qoft7}
\end{equation}
where
\begin{equation}
H(\Gamma;u) = \int_u^\infty d v_\perp f(v_\perp) \int d \gamma \, d b\, b 
G\biggl(\gamma;{v_\perp \Delta t\over b}\biggr)
\delta\biggl[\Gamma - \biggl({4 G M \over \alpha_0 b c^2}\biggr)^{-2}\gamma
\biggr].
\label{eqn:Hdef2}
\end{equation}
Carrying through the derivation, one obtains the analog of equation
(\ref{eqn:totaltime}),
\begin{equation}
T_{\rm tot} = {1\over 3}{\cal N}^3
\biggl({4 G \rho N \langle v_\perp^2\rangle \Delta t\over 
\mu_{\rm min}\alpha_0 c^2}\biggr)^{-2}
\gamma_* T_0,
\label{eqn:totaltime3}
\end{equation}
where $\langle v_\perp^2\rangle$ is the mean square transverse speed.
That is, equations (\ref{eqn:totaltime}) and (\ref{eqn:totaltime3}) 
are identical except 
$r_{\rm max}v_*\rightarrow \langle v_\perp^2\rangle/\mu_{\rm min}$.
 For a Gaussian, $\langle v_\perp^2\rangle=2\sigma^2$, so this relation
can be written
\begin{equation}
r_{\rm max} \rightarrow {4\over \pi}\,{v_*\over\mu_{\rm min}}
= 90\,{\rm pc}\,{v_*\over 33\,\kms}\,
\biggl({\mu_{\rm min}\over 100\,{\rm mas}\,{\rm yr}}\biggr)^{-1}.
\label{eqn:mursub}
\end{equation}
I find that with this substitution, the curves produced by equations
(\ref{eqn:qoft7}) and (\ref{eqn:Hdef2}) are almost identical to those
produced by equations (\ref{eqn:qoft4}) and (\ref{eqn:Hdef})
except that the cutoffs are increased by a factor 1.2.  Thus,
proper-motion selection and distance selection produce essentially the
same results, provided they are converted using equation (\ref{eqn:mursub}).

\section{Identification of Pairs}

	From equation (\ref{eqn:totaltime2}), the total observing time
required to measure the mass of a fixed number of stars declines as
$r_{\rm max}^{-2}$.  From Figure \ref{fig:two}, one sees that including
the magnitude bins $V_s=18,19$ is roughly equivalent to increasing
the total number of sources $N$ by a factor 2, and from equation 
(\ref{eqn:totaltime2}), the observing time is reduced as $N^{-2}\sim 0.25$.
This estimate is confirmed by the offset between the two bold curves
in Figure \ref{fig:two}.

	Hence, if it were possible to push out to fainter sources and
more distant lenses, it would certainly be profitable to do so.  I therefore
now investigate the constraints governing the identification of lens-source
pairs.  The basic problem is that if these pairs are to be close enough
for astrometric microlensing in, say, 2008, then they are {\it already} very
close.  That is, their separation $\Delta\theta_{\rm now}$ is 
\begin{equation}
\Delta\theta_{\rm now} = 0.\hskip-2pt ''5\,{v_\perp\over 30\,\kms}\,
{T_{\rm event}-T_{\rm now}\over 8\,\rm yrs}\,
\biggl({r\over 100\,\rm pc}\biggr)^{-1}.
\label{eqn:panic}
\end{equation}
Thus,  it would be difficult to conduct a large-scale survey
for such pairs.

	Fortunately, the US Naval Observatory is soon planning to release
an all-sky proper motion survey, USNO-B 
(D.\ Monet 1999, private communication).  Even so, the identification of
pairs is not trivial, and becomes more difficult both for fainter sources
and larger lens distances (or lower proper motions).  

\subsection{Unblemished Survey Data}

	I begin by considering the problem of candidate pair identification 
when the proper-motion survey data conform to ``typical'' specifications.
In fact, the underlying data sets are heterogeneous, with substantially
longer baselines in the north than the south.  For definiteness, I will
consider proper motions derived from 2 epochs, one in 1955 and the other
in 1990.  This baseline is appropriate for the northern hemisphere.
The anticipated proper-motion error is $4\,\rm mas\,yr^{-1}$,
corresponding to 100 mas errors in each position measurement.  This implies
an error of about 120 mas in the predicted positions of the source and
the lens in 2008, or 170 mas error in their relative position.  (Generally, 
only the error in one direction -- that of the impact parameter -- comes
into play.)\ \ Is this good enough?  Let us suppose that all pairs requiring
$T<T_{\rm max} = 200\,$hours are to be observed. From equation (\ref{eqn:tmax})
this corresponds to ${\cal N}\sim 6$.  For $\gamma=10$,
equation (\ref{eqn:taudef}) implies $\alpha \leq
115\,\mu$as $10^{0.2(V_s-17)}$, 
and so from equation (\ref{eqn:alphaeval})
\begin{equation}
\beta_{\rm max} = 
700\,{\rm mas}\,{M\over M_\odot}\,\biggl({r\over 100\,\rm pc}
\biggr)^{-1}\,10^{0.2(17-V_s)}.
\label{eqn:betacomp}
\end{equation}
Hence, for $M\sim M_\odot$, $\beta_{\rm max}$ is greater than $170\,$mas
even for $r=200\,$pc {\it or} $V_s=18$.  Thus while it would still be necessary
to do additional astrometry in order to prepare for the observations,
relatively few candidates would be rejected by this astrometry.  By contrast,
for $M=0.2\,M_\odot$, $r=200\,$pc, and $V_s=18$, $\beta_{\rm max}\sim 30$ mas.
In this case, it would be necessary to examine about 6 candidates drawn
from the proper-motion survey to find one suitable for a mass measurement.

	In the southern hemisphere, the baselines are shorter, so the 
proper-motion errors are about twice as big.  Hence, about twice as many
candidates need to be examined.

\subsection{Compromised second epoch}

	If the time of closest approach is 2008, then at the time of the 
second epoch of the proper motion survey, say 1990, a typical foreground
star will be
separated from the line of sight to the source by of order 100 AU.  This
corresponds to $1''$ at $r=100\,$pc.  As discussed in \S\ 3.1 in the
analysis of Figure \ref{fig:one}, lens candidates tend to be moving faster
than the population as a whole, so the actual typical separation will be
somewhat larger.  Nevertheless, these values are close to the resolution
limit of the surveys and become even less favorable at greater distances.
In addition, bright lens stars will entirely blot out a substantial region
around them on the survey plates, 
preventing the detection of candidate source stars at all.
For example, I find that on the Palomer Observatory Sky Survey (POSS),
$V=8$ stars (the approximate completeness limit of the {\it Hipparcos}
catalog) tend to blot out a region with a $10''$ radius.

	However, even the complete loss of the second-epoch positions of
candidate sources is not crippling.  The proper motion of the bright lens
candidate can still be measured, and its position in 2008 predicted.
This region can then be examined on the first epoch plates for potential
candidates (assuming that the lens is not bright enough to have blotted out
this region even at this earlier epoch).  Of course, in the intervening
$\sim 50$ years, these candidates will have moved, but probably not by
very much.  For example, at high latitudes $(|b|\ga 20^\circ)$, disk
sources typically lie at 3 disk scale heights or about $|\csc b|\,$kpc.
Hence, in 50 years, they will typically move $300 |\sin b|$ mas which even
at $b=90^\circ$ is not much more than the error in the expected position
(see \S\ 4.1).  Closer to the plane, the motion will be even less.  Thus,
without a second epoch, more candidates will have to be examined at
high latitudes (but these contain a minority of the candidates anyway) and
there will be hardly any effect at low latitudes.

\subsection{What can be done now?}

	The USNO-B catalog has not yet been released.  However,
it is still possible to begin the search for candidates using the
Luyton (1979) proper-motion catalog in combination with the USNO-A2.0
all sky astrometric catalog (Monet 1998).  The conditions of such a search
are fairly well described by \S\ 4.2.  

	Based on a comparison to Hipparcos data, I.N.\ Reid 
(1999 private communication) estimates that the Luyton (1979) proper motions 
are typically accurate to about 10 mas yr$^{-1}$, and that the positions
are accurate to a few arcseconds.  By identifying the Luyton (1979) stars
in the USNO-A2.0 survey, one could fix the $\sim 1955$ positions to 
$\sim 100\,\rm mas$, and hence the 2008 positions to $\sim 500\,\rm mas$.
One could then search the USNO-A2.0 catalog for background stars whose
$\sim 1955$ positions lay along the 2008 path of the Luyton star.  As 
discussed in \S\ 4.2, these stars could be expected to move about 
$300 |\sin b|$ mas which is generally small compared to the uncertainty in
the position of the foreground star.  Hence, 2008 source-lens separations could
be predicted to $\sim 0.\hskip-2pt ''5$.  For pairs that were sufficiently
close, the separation could be measured on the POSS II plates to refine
the prediction.  Follow-up observation could then be made of 
those pairs surviving this test.

\section{Stellar Halo Lenses}

	Halo stars are about 500 times less common than disk stars
(Gould, Flynn, \& Bahcall 1998), i.e., $\rho\sim 6\times 10^{-5}\,M_\odot
\,\rm pc^{-3}$, and they are typically moving about 5 times faster.  Let
us suppose that they could be spotted to $r_{\rm max}=1\,$kpc (see below),
and let us take $N=4\times 10^8$ in accord with the discussion of
Figure \ref{fig:two}.  Then,
from equation (\ref{eqn:totaltime2}), it would be possible to obtain the
masses of 5 halo stars with about 150 hours of observation.

	At $r_{\rm max}=1\,$kpc, it is still appropriate to approximate
the density of the stellar halo as uniform.  However, it is no longer
appropriate to treat the sources as being infinitely far away.  As
mentioned in \S\ 4.2, at $b=90^\circ$, typical disk sources are at 1 kpc.
However, the disk sources are farther away at lower latitudes
where there are the greatest fraction of sources in any case.  Hence, given
the level of approximation of the present study, I will ignore this
modest correction.

	Of course, the radius within 1 kpc contains of order 300 times
more stars than the radius within 100 pc, so identifying a relatively 
complete sample of halo stars seems like a formidable task at first sight.
In fact, for stars of fixed color (and approximated as black bodies -- or
at least as deviating from black bodies in similar ways), we have
\begin{equation}
t_{\rm cross} = k_{(B-R)_0} 10^{-0.2 R_0}\mu^{-1},
\label{eqn:tcross}
\end{equation}
where $k_{(B-R)_0}$ is a constant that depends on the dereddened
color, $R_0$ is the dereddened magnitude, and $t_{\rm cross}$ is the time
it takes the star to cross its own radius.  For halo stars, 
$t_{\rm cross}\sim 10^3\,$s, which is significantly different from the
value for other common classes, 
$2\times 10^2\,$s for disk white dwarfs,
$5\times 10^3\,$s for thick-disk stars, and
$2\times 10^4\,$s for main sequence stars.  Thus it should not be difficult
to find halo star candidates in a proper-motion catalog with colors.  
The sample will be somewhat contaminated with fast-moving thick-disk
stars, but these are also of considerable interest because of their
low metallicity.

	A more significant problem is that if the survey is limited to
$V\sim 20$, then at 1 kpc, the bottom of the luminosity function
$M_V>10$ is not detectable.  These fainter stars contain about half the
spheroid mass, implying that the above estimate of the observation time
required to measure 5 masses should be multiplied by 4 to about 600 hours.
Of course, if one were willing to settle for 3\% measurements in place
of 1\%, this estimate would come down by an order of magnitude.

\begin{equation}
\label{eqn:}
\end{equation}

{\bf Acknowledgements}: I thank Scott Gaudi and Samir Salim for their 
careful comments on the manuscript, and Neill Reid for his helpful
suggestions regarding the Luyton catalog.  
This research was supported in part by grant AST 97-27520 from the NSF.

\newpage

\begin{figure}
\caption[junk]{\label{fig:one}
Total number of mass measurements ${\cal N}$ 
as a function of total required observing time $T_{\rm tot}$.  The solid
curve is for $M=0.1\,M_\odot$ and the bold curve is for $M=M_\odot$.
The dashed line is the analytic approximation (but without the cutoff)
given by eq.\ (\ref{eqn:totaltime2}).
}
\end{figure}

\begin{figure}
\caption[junk]{\label{fig:two}
Total number of mass measurements ${\cal N}$ 
as a function of total required observing time $T_{\rm tot}$, for
mass $M=M_\odot$.
The $V_s=17$ curve (same as in Fig. \ref{fig:one}) is shown as a
bold dashed line, and the others $V_s=14,15,16,18,19$ are shown as solid
lines.  The curves can be separately identified by noting that the
cutoff increases with magnitude.
The upper bold line shows the result of combining all of these bins
($14\leq V_s\leq 19$) while
the lower bold line shows the result of combining the four brightest bins
($14\leq V_s\leq 17$).
}
\end{figure}

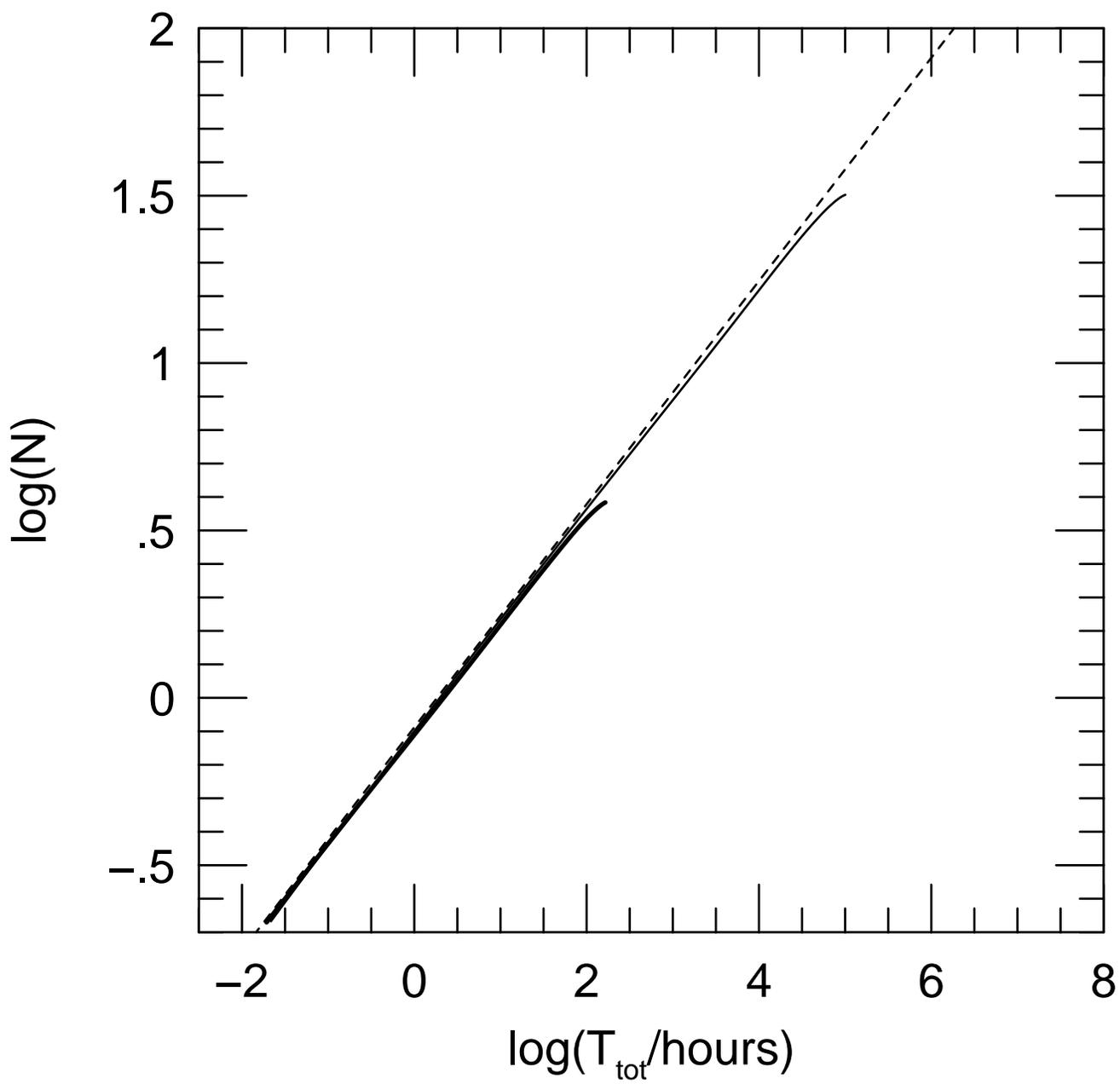
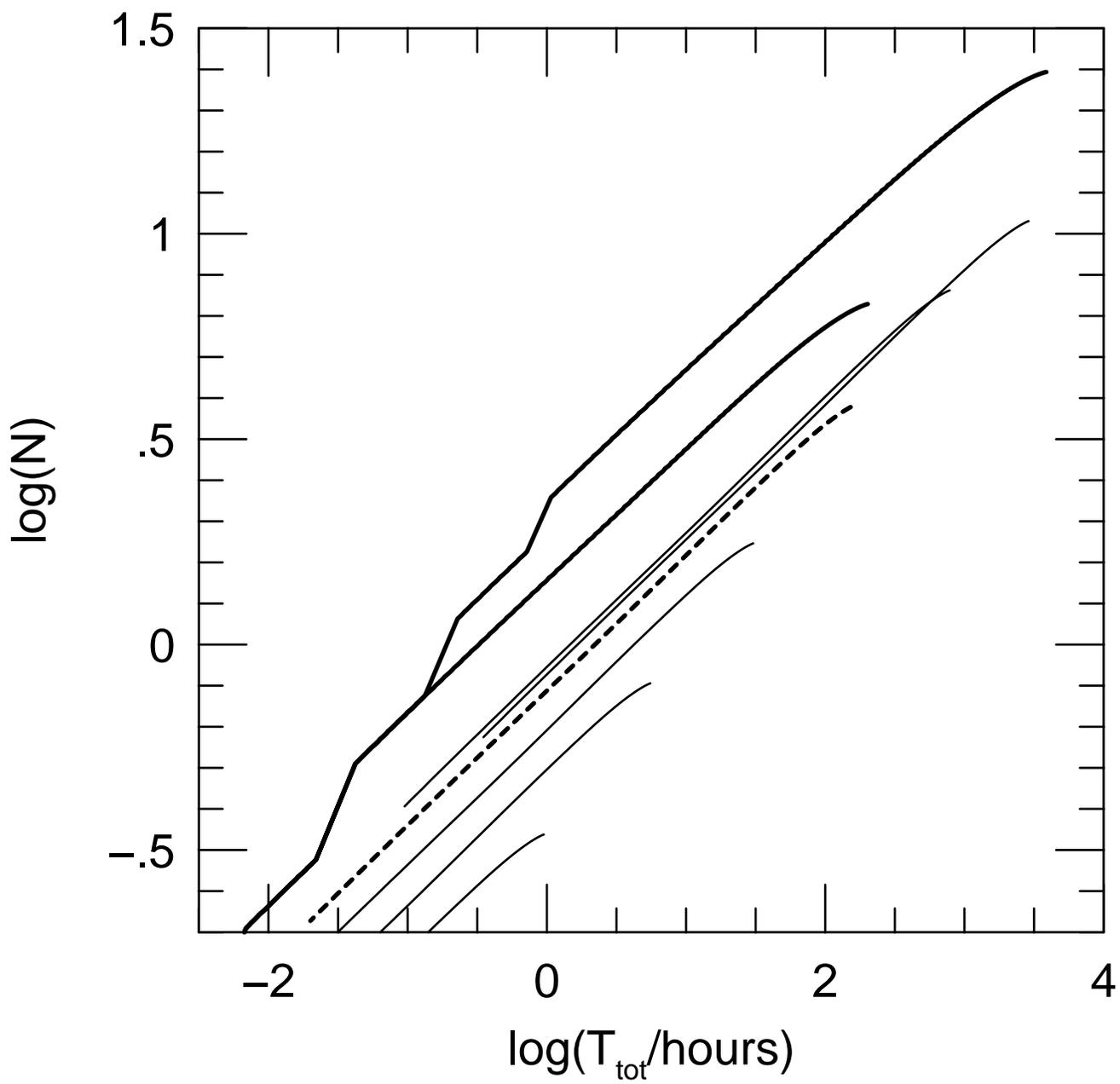
\end{document}